\documentclass[prl,aps,twocolumn,groupedaddress,floats,final,nopacks,superscriptaddress]{revtex4-2}
\usepackage{graphicx}
\usepackage{subfigure}
\usepackage{amssymb}
\usepackage{latexsym}
\usepackage{epsfig}
\usepackage{psfrag}
\usepackage{dcolumn}
\usepackage{amsmath,amsfonts,amssymb,bm,ulem}
\usepackage{color}
\usepackage{float}

\newcommand{\be}{\begin{equation}}
\newcommand{\ee}{\end{equation}}
\newcommand{\bea}{\begin{eqnarray}}
\newcommand{\eea}{\end{eqnarray}}

\usepackage[colorlinks=true, linkcolor=blue, citecolor=blue, urlcolor=blue]{hyperref}

\begin{document}
\title{Numerically Exact Study of Flat-Band Superconductivity}
\author{I.S. Tupitsyn}
\email{itupitsyn@physics.umass.edu}
\affiliation{Department of Physics, University of Massachusetts, Amherst, MA 01003, USA}
\author{B. Currie}
\email{benjamin.d.currie@kcl.ac.uk}
\affiliation{Physics Department, King's College London, Strand, London WC2R 2LS, UK}
\author{B.V. Svistunov}
\affiliation{Department of Physics, University of Massachusetts, Amherst, MA 01003, USA}
\affiliation{Physics Department, King's College London, Strand, London WC2R 2LS, UK}
\author{E. Kozik}
\affiliation{Physics Department, King's College London, Strand, London WC2R 2LS, UK}
\author{N.V. Prokof'ev}
\affiliation{Department of Physics, University of Massachusetts, Amherst, MA 01003, USA}
\affiliation{Physics Department, King's College London, Strand, London WC2R 2LS, UK}



\begin{abstract}

Current theories of high-temperature superconductivity in flat-band systems predict a linear dependence of the transition temperature on the attractive interaction, $T_c(U) = c|U|$.
However, the value of $c$ and the full nonlinear $T_c(U)$ curve---with a maximum at large $|U|$---are known beyond mean-field and quantum geometry estimates only for systems with isolated flat bands.
Using a controlled diagrammatic Monte Carlo technique, we trace the onset of superfluid response in the Lieb lattice with attractive Hubbard interaction. Focusing on the half-filled flat-band case, where the ordering mechanism differs fundamentally from both BCS and BEC (preformed Cooper pair) scenarios, we find that the pairing response diverges linearly with decreasing temperature over a broad range of $U$, leading to a sharp crossover to long-range correlations at a characteristic temperature $T_*$, which provides a controlled upper bound on $T_c$. The highest $T_*$ occurs when all three bands touch at a single momentum point, potentially corresponding to high $T_c$ values.

\end{abstract}

\maketitle

\noindent \textit{Introduction.} Flat-band superconductivity (FBSC) in multi-band systems is an extremely counter-intuitive concept. On the one hand, the starting point is a system of ideal fermions with infinite bare mass $m_0=\infty$, which cannot support current-carrying states. On the other hand, all energy scales characterising interaction-induced properties, including the critical temperature of the superfluid transition, $T_c$, are expected to scale linearly with the interaction $U$ in the $U \to - 0\,$ limit (this is also the prediction of the mean-field and quantum geometry considerations \cite{Khodel1990,Kopnin2011,Heikkila2011,PT2015,PT2016,PT2017,Iskin2019}),
implying the possibility of achieving high-temperature superconductivity with
\be
T_c \, =\,  c \, |U| \, ,
\label{TcU}
\ee
where $c$ is a certain dimensionless constant \cite{note8}.

Key ideas for overcoming the $m_0=\infty$ limitation in interacting systems were developed in Refs.~\cite{PT2015,PT2016,PT2017}, which showed that the quantum metric---the real part of the quantum geometric tensor characterising Hilbert-space geometry (see, e.g., Refs.~\cite{QGT-Rev,Yu2025})---enables Cooper pair propagation in multi-band systems, yielding a nonzero superfluid weight. Subsequent works examined the dependence of superconducting properties on the number of bands~\cite{PT2022-1,PT2022-2,multi2022,PT2025}. Formally, representing the Hubbard interaction as a matrix in the two-body band basis reveals non-zero off-diagonal elements that generate dispersive single-particle states, mobile bound pairs, and hence superconductivity. Their strength and role depend on the lattice structure, the number of bands, and the inter-band gaps~\cite{PT2022-1,PT2022-2,multi2022,PT2025}.
So far, numerically exact studies of the problem were carried
out only for isolated flat-band systems with suppressed
inter-band transitions~\cite{Berg2020,Berg2023}.

Alongside the conceptual interest in FBSC as a non-BCS/non-BEC pairing scenario, there is an important quantitative aspect. Relation (\ref{TcU}) implies that even for small $|U|$, the critical temperature can be much higher than in conventional BCS superconductors, where $T_c$ is exponentially suppressed in the inverse coupling, provided the dimensionless coefficient $c$ is not anomalously small for purely numerical reasons. Addressing this quantitatively in a controlled manner is challenging, as Eq.~(\ref{TcU}) reflects the essentially non-perturbative nature of the problem arising from the macroscopic degeneracy of the underlying non-interacting ground state.

In this Letter, we present a controlled first-principles approach to the FBSC problem based on the Diagrammatic Monte Carlo (DiagMC) method. Although a perturbative expansion about a macroscopically degenerate flat band may appear ill-defined, a finite temperature $T$ introduces an additional energy scale and ensures a non-zero convergence radius and thus convergence in the regime $|U|/T \ll 1$. The essentially non-perturbative regime beyond the convergence radius can then be accessed by controlled series resummation, as recently demonstrated in the DiagMC description of the fractional quantum Hall state in the lowest Landau level~\cite{Currie2024}.

\noindent \textit{ Model.} As a paradigmatic example of FBSC, we consider the attractive Hubbard model on the Lieb lattice~\cite{Lieb1989} with the Hamiltonian
\be
H \, =\,  - \sum_{<mm'> \sigma} t_{m m'} \hat{c}^{\dag}_{m\sigma} \hat{c}^{\,}_{m'\sigma}
\, +\, U  \sum_{m}  \hat{n}_{m\uparrow} \hat{n}_{m\downarrow}\, .
\label{LH}
\ee
Here $m=\{\alpha, j\}$ is a combined (sublattice, unit cell) index, $<\ldots>$ symbol restricts summation over the pairs of nearest-neighbor lattice sites, $\hat{c}^{\dag}_{m \sigma}$ creates a fermion with spin projection $\sigma \in \{\uparrow,\downarrow\}$ on sublattice $\alpha=\{A,B,C\}$ within the $j$-th unit cell, $\hat{n}_{m \sigma}=\hat{c}^{\dag}_{m\sigma} \hat{c}^{\,}_{m\sigma}$, and $U<0$ is a sublattice-independent attractive interaction. The fermion density is fixed at three particles per unit cell, which corresponds to a half-filled flat band.

\begin{figure}[t]
\begin{center}
\includegraphics[width=1.0\linewidth] {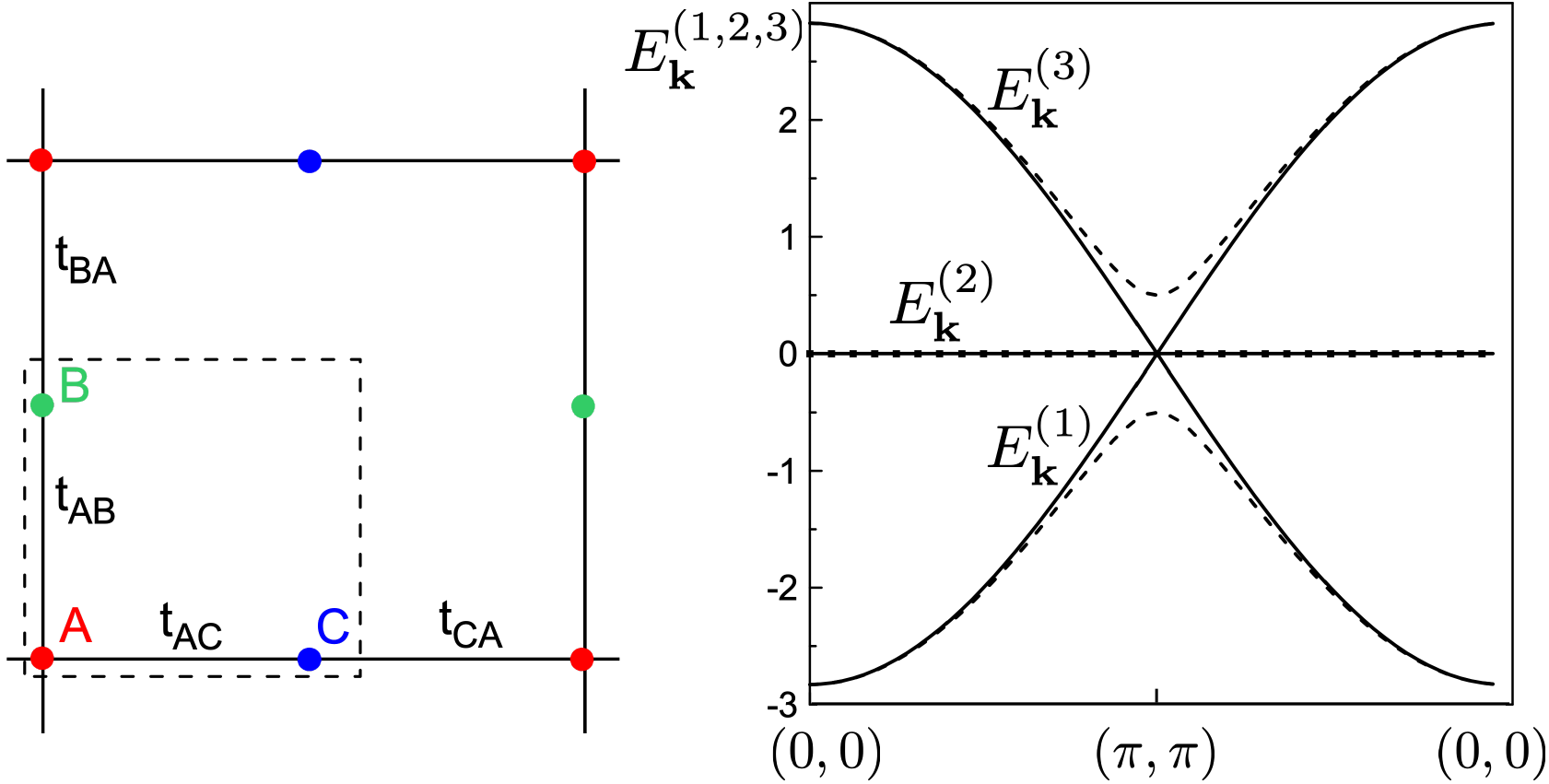}
\end{center}
\vspace{-5mm}
\caption{
Left: Lieb lattice with three atoms $\{ A, B, C \} $ in the unit cell and hopping amplitudes $t_{\alpha \beta}$ between the n.n. atoms. 
Right: Bare dispersions $E^{(1,2,3)}_{\mathbf{k}}$
of the three bands along the $k_x = k_y$ direction for all three setups described in the text: solid lines correspond to the standard Lieb lattice (case (i)), dashed (and dotted) lines correspond to the cases (ii) and (iii), which have identical dispersions (dotted lines mark flat bands)
}
\label{Fig1}
\end{figure}

We consider three flat-band setups that differ by the values of hopping matrix elements between the nearest-neighbor atoms, see the left panel of Fig.~\ref{Fig1}: (i) standard Lieb lattice with all bands touching at the momentum point $M = (\pi, \pi)$ in the Brillouin zone (BZ) (the lattice spacing is set equal to unity) and two lattices with $C_4$ symmetry broken down to (ii) $\sigma_h$ (reflection across $\hat{x}$ axis) and (iii) $\sigma_d$ (reflection across $\hat{x}=\hat{y}$ axis) with bands separated by the same gap. The corresponding bare dispersion relations are shown in the right panel of Fig.~\ref{Fig1}.

In the standard Lieb-lattice Hamiltonian, all nearest-neighbor hopping amplitudes are equal:  $t_{AB}=t_{BA}=t_{AC}=t_{CA}=t$ (we use $t$ as the unit of energy).
In setup (ii),  $t_{AB}=t_{BA}=t$, $t_{AC}=t(1-\delta_2)$, $t_{CA}=t(1+\delta_2)$ with $\delta_2=1/4$.
In setup (iii), $t_{AB}=t_{AC}=t(1+\delta_1)$, $t_{BA}=t_{CA}=t(1-\delta_1)$ with $\delta_1=1/(4\sqrt{2})$.
The central band is flat in all three (bare) cases. Setups (ii) and (iii) have band structures with the same band gap $\Delta = 0.5 t$, see Fig.~\ref{Fig1}, in order to explore the effects of symmetry breaking in different realizations of the Lieb lattice.


To detect the onset of superfluidity we compute the s-wave Cooper pairing susceptibility
\be
\chi \, =\,  \sum_{\alpha,\alpha', j} \int_0^\beta d\tau P_{\alpha \alpha'}(j, \tau) \, ,
\label{chi}
\ee
defined in terms of the pair correlator
\be
P_{\alpha \alpha'}(j, \tau) =   \langle c_{m' \uparrow}(\tau) \, c_{m' \downarrow}(\tau) \, c^{\dagger}_{m \uparrow}(0) \, c^{\dagger}_{m \downarrow}(0) \rangle
\label{correlaor}
\ee
with $m=\{\alpha, 0\}$ and $m'=\{\alpha', j\}$.
We do not differentiate between the irreducible representations
by assuming that for on-site $U<0$ the leading s-wave channel is captured by $P$, and, thus, $\chi$ diverges at $T\to T_c+0$. 
For computational convenience, we subtract the non-interacting contribution, $\chi_0$, based on the product of bare Green's functions, and use it for data normalization:
\be
I = \chi_0/[\chi-\chi_0] \, .
\label{I_SC}
\ee
In what follows we refer to $I$ as ``inverse susceptibility".

The Berezinskii-Kosterlitz-Thouless theory \cite{KT} predicts the following asymptotic form of $I$ near the critical point:
\be
I(T) \, \propto \, e^{-b /\sqrt{T-T_c}} \qquad \qquad (T \to T_c + 0 )  \, .
\label{chi_KT}
\ee
Here  $b$ is a non-universal constant controlled by the vortex core energy and associated with the density of vortex–antivortex pairs.
On approach to the BKT point, this law follows from the exponential divergence of correlation length~\cite{KT}, which cuts off summation at long distances in (\ref{chi}).
For vortex energy much larger than $T_c$ the universal regime (\ref{chi_KT}) is exponentially difficult to resolve, not only numerically but also experimentally. In this situation, rather than revealing the true critical temperature, $I$ will drop to near zero at a crossover temperature $T_* > T_c$, which signals the onset
of long-range superfluid response and does not need to be close to $T_c$. Experimentally, a sharp drop of $I(T)$ on approach to $T_*$ leads to two physical effects irrespective of $T_*$ proximity to $T_c$: a sharp drop in resistivity across $T_*$ and a genuine superfluid transition at $T_c \approx T_*$ in a 3D system consisting of weakly coupled 2D layers.


\begin{figure}[t]
\centering
\includegraphics[width=0.9\linewidth] {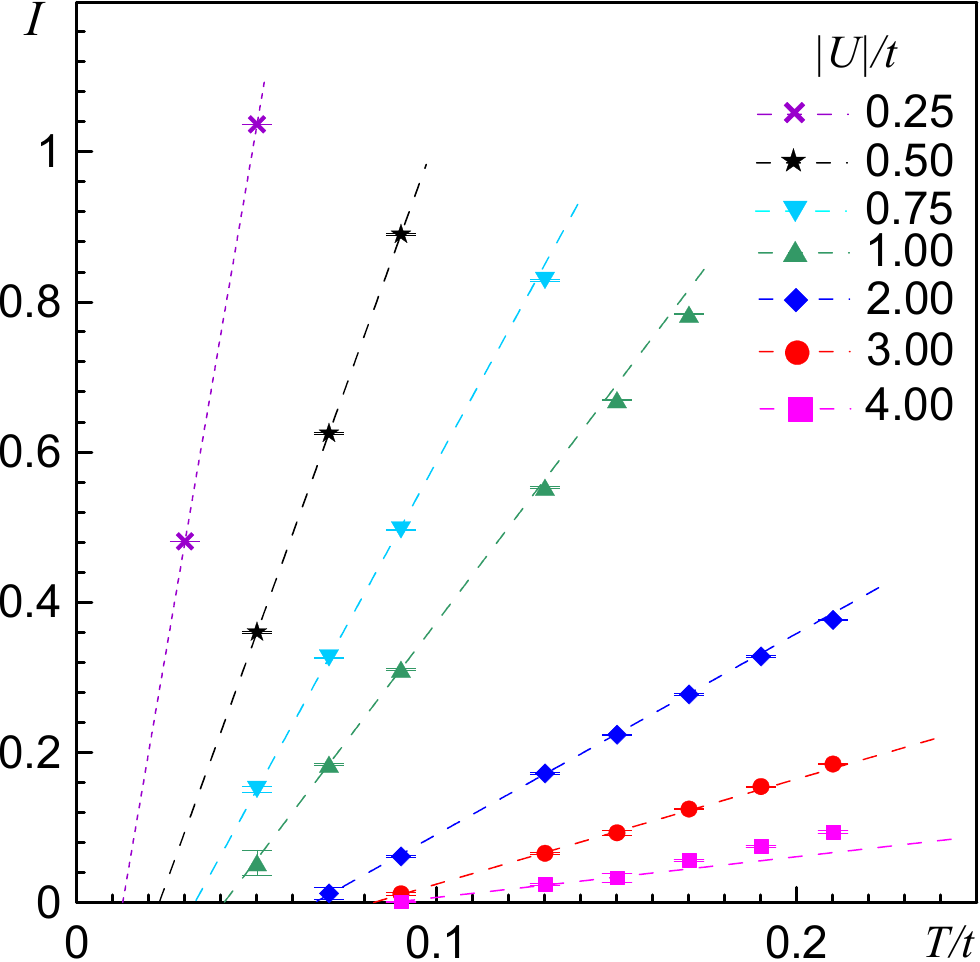}
\caption{
Inverse susceptibility $I$ of the standard Lieb lattice as a function of temperature for various values of $U$. Extrapolation towards $I=0$ yields the temperature $T_* > T_c$ at which a sharp crossover towards a dramatic enhancement of the pairing response takes place. [For setup (ii) data see End Matter subsection "Inverse paring susceptibility $I(T)$".]
}
\label{fig:straight}
\end{figure}

\begin{figure}[t]
\centering
\includegraphics[width=0.9\linewidth] {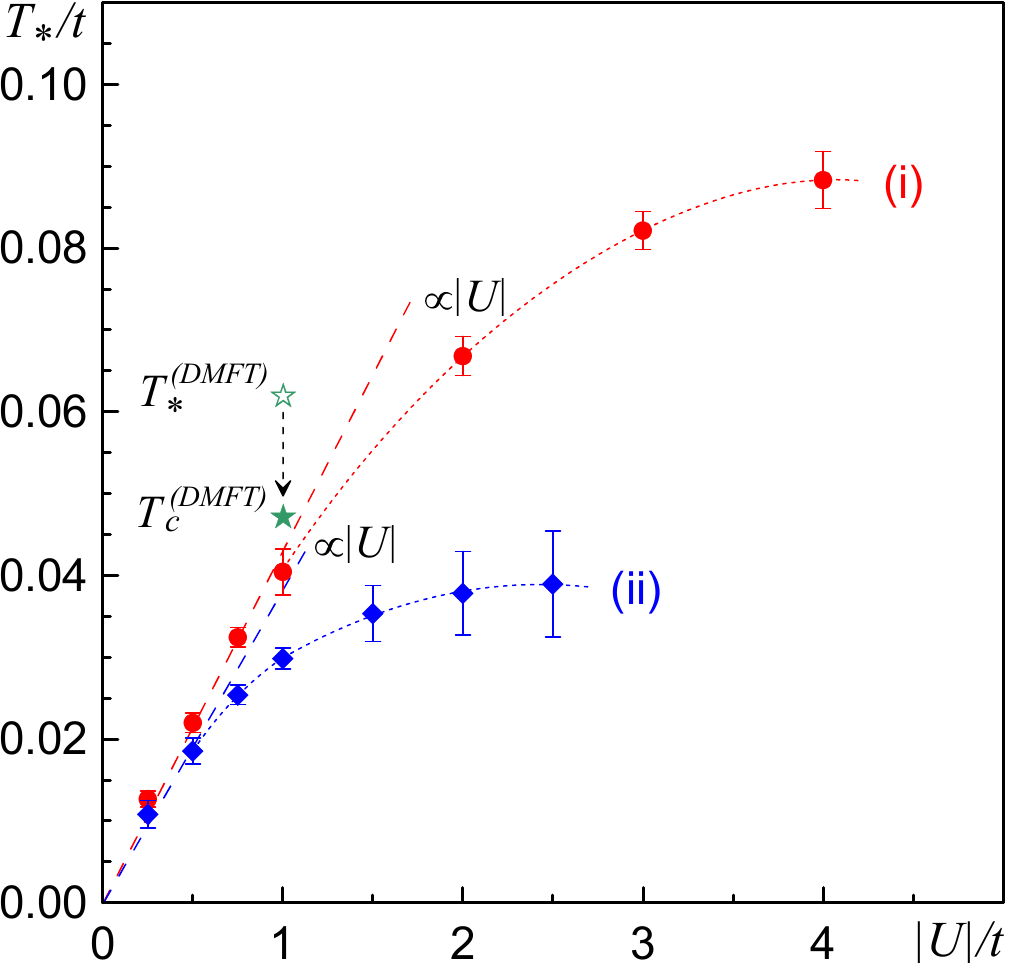}
\caption{Crossover temperature as a function of $U$ for two Lieb lattice setups: (i) - red circles, (ii) - blue diamonds. Red and blue dashed lines are the linear $T_* \propto |U|$ laws. By empty green star we show the recent DMFT result \cite{PT2025} for the mean-field transition temperature; DMFT data for superfluid stiffness~\cite{PT2025} allow one to predicting $T_c$ using the Nelson-Kosterlitz criterion  (filled green star).
}
\label{fig:T_star}
\end{figure}

\noindent \textit{Methods.} In the DiagMC framework~\cite{Prokofev2007, VanHoucke2010, Kozik2010}, $\chi$ is represented as an expansion in powers of a certain Hamiltonian (or effective action) parameter; we consider expansion in powers
of the bare coupling, $\chi(U)=\sum a_n U^n$, with the bare Hartree diagrams absorbed into the chemical potential via appropriate shifts~\cite{VanHoucke2010}. The coefficients $a_n$ are computed numerically exactly as a sum of all connected Feynman diagrams of order $n$. The only source of controlled systematic error is truncation of the expansion at some sufficiently large order $N$. To compute $a_n$ with small statistical errors up to $N\sim 8$, we employ the recently developed combinatorial summation (CoS) algorithm \cite{Kozik2024}, which deterministically sums all $\propto (n!)^2$ diagram integrands using a computational graph. Sign cancellations between integrands substantially reduce the MC variance of the subsequent integration over internal space-time coordinates; see Ref.~\cite{Kozik2024} for details. The final answer for $\chi(U)$ is reconstructed using the standard Dlog Pad\'e~\cite{Pade1961} and Integral Approximant methods~\cite{Hunter1979IA}. The spread among different approximants provides an estimate of the systematic extrapolation error, as described in Ref.~\cite{PadeSK2019} and illustrated for our calculation of $\chi$ in End Matter. The DiagMC error bars in Figs.~\ref{fig:straight} and \ref{fig:T_star} include both the statistical uncertainties in $a_n$ and the extrapolation error.

One might ask to what extent advanced but lower-cost diagrammatic schemes that account for momentum-dependent correlations can capture the FBSC physics quantitatively and be used for exploration of the broad parameter space. To this end, we developed an extension of the four-channel self-consistent diagrammatic theory, the Bold4 approach~\cite{MagCor2017}, that incorporates leading vertex corrections and generalizes it to multi-band systems (see ``End Matter" for details). This, Bold4+, technique accounts for all diagrammatic contributions up to order $U^4$, is asymptotically exact in the low-$U$--high-$T$ limit, and was used to cross-validate the CoS calculations in this regime.

\noindent \textit{Key results.}
Our calculations at small-to-moderate $U$ reveal behavior markedly different from the asymptotic form (\ref{chi_KT}) over a broad temperature range where $\chi$ grows strongly: Fig.~\ref{fig:straight} shows an almost perfectly linear $I(T)$, a hallmark of Gaussian criticality. Deviations from linearity--possibly signalling a crossover toward BKT behavior--are only weakly visible at $|U| = 4t$. By itself, the linear dependence does not point to any specific pairing scenario and instead indicates the absence of pronounced nonlinear bosonic field effects at $|U|\lesssim 3$, which underpin the crossover to BKT criticality. Therefore, for this system, $T_c$ cannot be reliably extracted from numerical data for $I(T)$. Instead, we focus on the characteristic temperature $T_*$, defined by linear extrapolation $I(T) \propto T-T_*$.

Following Ref.~\cite{Lenihan2022}, $T_c$ can in principle be extracted from the divergence of the diagrammatic expansion for $I(U)$ at the singularity $T_c(U)$ [Eq.~\eqref{chi_KT}]. However, at our maximum expansion order of 8, the BKT form~\eqref{chi_KT} does not fit the series coefficients better than a generic power-law singularity. Consequently, the extracted singularity corresponds to a variant of $T_*$--consistent with that obtained from extrapolating $I(T)$--rather than the BKT $T_c$.


Our results for  $T_*$ as a function of $U$ are shown in Fig.~\ref{fig:T_star}.
Basic dimensional analysis leading to the result (\ref{TcU}) yields a similar result for $T_*$:
\be
T_* \, =\,  c_* \, |U| \qquad \qquad (U \, \to \, -0) \, .
\label{T_star_of_U}
\ee
As $T_*$ is not necessarily close to $T_c$, one cannot assume \textit{a priori} that $c_*$ will be close to $c$; only the inequality $c < c_*$ can be guaranteed.

\begin{figure}[t]
\begin{center}
\includegraphics[width=0.9\linewidth]{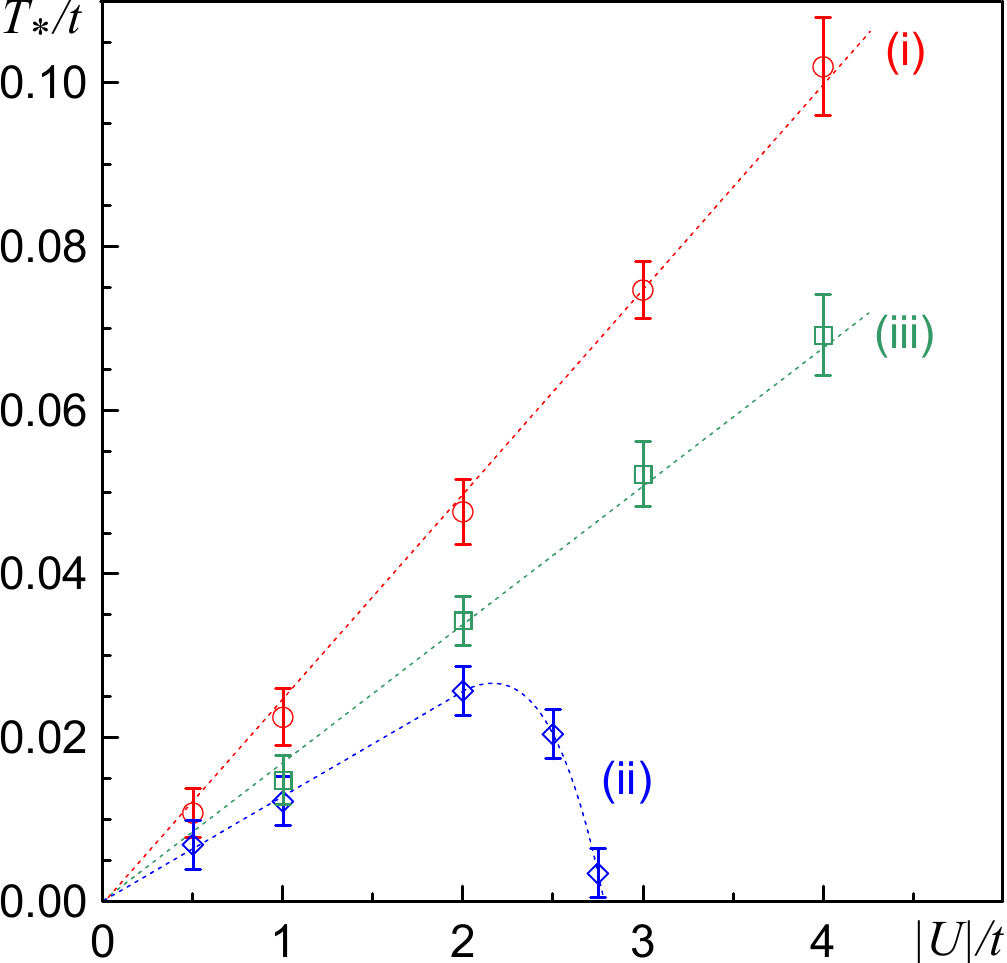}
\end{center}
\vspace{-5mm}
\caption{
$T_*(U)$ within the  Bold4+ scheme for all three setups: (i) - red circles, (ii) - blue diamonds, (iii) - green squares.
}
\label{Fig4}
\end{figure}

We find that for the standard Lieb lattice--setup (i)--this linear scaling persists up to $|U| \sim t$ (see Fig.~\ref{fig:T_star}) with $c_*=0.042(5)$. At $|U| \gtrsim t$, the value of $T_*$ starts levelling off, reaching a maximum of $\approx 0.09t$ at $|U| \approx 4t$. For setup (ii), the maximum of the $T_*(U)$ curve is significantly lower and takes place at smaller $|U|/t \approx 2.5$; we expect that $T_*$ will decrease with interaction at $|U|/t \gtrsim 2.5$.
Extending the parameter range beyond that considered in Fig.~\ref{fig:T_star} would require simulations at substantially lower temperatures (for smaller $|U|$) or higher expansion orders (for larger $|U|$), without altering our main conclusions. [We also note that estimating $U_*(T)$ for fixed $T$ from the vanishing of $I$ as a function of $U$ gives consistent curves to those shown in Fig.~\ref{fig:T_star}.]
Computationally demanding high-order CoS simulations were not performed for setup (iii): on the basis of Bold4+ simulation results shown in Fig.~\ref{Fig4} we expect that $T_*(U)$ curve for setup (iii) falls in between the corresponding curves for setups (i) and (ii).
%

\textit{Discussion}. Linear dependence of $I(T)$, while indicating the irrelevance of vortices down to $T=T_*$, does not necessarily imply that the numerically exact results of Fig.~\ref{fig:T_star} can be accurately captured--even at small $U$--by mean-field or perturbative theories. It is therefore instructive to compare the results of Fig.~\ref{fig:T_star} with predictions of such theories to assess their quantitative accuracy. From a practical standpoint, an accurate approximate theory would be highly valuable, given the substantial computational cost of controlling error bars by reaching high expansion orders in DiagMC-CoS. Recent dynamical mean-field theory (DMFT) calculations \cite{PT2025} predicted that the mean-field transition temperature $T^{\mathrm{(DMFT)}}_*$--an analog of our $T_*$--follows Eq.~(\ref{T_star_of_U}) with $c^{\mathrm{(DMFT)}}_*=0.062$ for the standard Lieb lattice. However, DMFT allows one to compute the superfluid stiffness and this was used to obtain the transition temperature via the universal Nelson-Kosterlitz relation \cite{NK} with the result $T^{\mathrm{(DMFT)}}_c \approx 0.047 U$. Both $T^{\mathrm{(DMFT)}}_*$ and $T^{\mathrm{(DMFT)}}_c$ for $|U|=t$ are displayed in Fig.~\ref{fig:T_star}. Since our simulations have no access to the superfluid stiffness, the only physically relevant comparison is between our $T_*$ and $T^{\mathrm{(DMFT)}}_*$.

We find that $I(T)$ curves obtained via Bold4+ also demonstrate an almost perfect linear scaling with temperature, see Fig.~\ref{I_T_data} in End Matter. However, the resulting $T_*(U)$ curves (Fig.~\ref{Fig4}) differ markedly from the controlled results in Fig.~\ref{fig:T_star}; specifically, the Bold4+ value of $c_* \approx 0.02$ for the standard Lieb lattice is a factor of two smaller than the DiagMC result. When the $C_4$ symmetry is broken in setup (ii), the initial linear $T_*(U)$ dependence breaks down at $|U| \approx 2t$ and plummets toward zero by $U \approx -2.75 t$. In this case, we observe notable interaction-induced dispersion in the central band, with a width $W_2$ that along the $\hat{x}=\hat{y}$ direction increases with $U$ up to $W_2\sim 0.1t$ at $U \sim -2.5t$. Simultaneously, the band gaps increase by $\sim 30\%$. Nonetheless, the qualitative distinction between the $T_*(U)$ curves in cases (i) and (ii) at moderate $|U|\lesssim 2t$ is similar to that observed in Fig.~\ref{fig:T_star}, with the standard Lieb lattice case consistently displaying higher values of $T_*(U)$.

Finally, we use the Bold4+ scheme to investigate the difference between the setups (ii) and (iii), which have the same bare band structure but different residual symmetry. Results for (iii) in Fig.~\ref{Fig4} are remarkably similar to (i) with linear scaling
extending up to $U \sim -4t$ but with a smaller $c_*$.
In both setups the central band acquires interaction-induced dispersion along the non-symmetric directions---along $\hat{x}$ and $\hat{x} = \pm\hat{y}$ in (ii) and along $\hat{x}$ and $\hat{y}$ in (iii)---and the band gaps increase with $|U|$.
It appears that breaking the $\sigma_{d}$ symmetry and, consequently,
equivalence between the $B$ and $C$ sites, on which the flat band state is based, is the most ``destructive" for FBSC effect. The mean-field $T_c(U)$ curves in Ref.~\cite{PT2022-1} are qualitatively similar to what we have for cases (i) and (ii) in Fig.~\ref{Fig4}, but they are above our upper-bound
$T_*(U)$ values in Fig.~\ref{fig:T_star}.

\noindent \textit{Conclusions.} Using the controlled diagrammatic Monte Carlo technique, we traced the onset of the superfluid response in a paradigmatic model for flat-band superconductivity: the half-filled Lieb lattice with attractive on-site interaction. Our data (Fig.~\ref{fig:straight}) demonstrate that the off-diagonal two-particle correlations developing with decreasing temperature exhibit a Gaussian character across a broad range of temperatures and interaction strengths. These correlations show a sharp crossover to long-range superfluid behavior at a specific temperature $T_*$ which, on the one hand, provides a rigorous upper bound on the critical temperature of the BKT transition in 2D and, on the other, carries a distinct physical meaning as the temperature at which (i) the system's resistivity drops dramatically and (ii) a realistic system of weakly coupled 2D layers would undergo a genuine superfluid transition with $T_c \approx T_*$.

We considered three characteristic cases: a gapless $C_4$-symmetric system and two gapped systems with $C_4$ broken down to $\sigma_h$ and $\sigma_d$. We computed $T_*$ dependence on the attractive Hubbard interaction $U$ and found that it is linear at small $|U|$ (see Fig.~\ref{fig:T_star}), but levels off and passes through a maximum at strong coupling. The largest proportionality coefficient $c_* = 0.042(5)$ and the highest value $T_*\sim 0.09t$ at its maximum at $|U| \sim 4t$ were observed for the standard gapless Lieb lattice. Band gaps and breaking the symmetry between the B and C atoms forming the flat-band state is detrimental for flat-band superconductivity.
Future work should explore alternative flat-band Hamiltonians and lattice geometries to identify the optimal material candidates for maximizing the transition temperature. Our finding that $T_*$ can reach values as high as $0.09 t$ implies that, under the right conditions and with typical hopping amplitudes of approximately $0.3-0.5~$eV, there are reasonable prospects for achieving high transition temperatures. In End Matter we show that attractive Hubbard interaction with
$3 \lesssim |U|/t \lesssim 4$
can be generated by moderate electron-phonon coupling. Promising flat-band Lieb-lattice systems include cold atoms in optical lattices, photonic lattices, STM patterned surfaces, synthesized 2D polymer layers, and predicted materials such as Ca$_2$I (the so called ``Lieb electride" \cite{Electride2025}).

\begin{acknowledgments}
\noindent \textit{Acknowledgments.} We thank R.P.S. Penttil\"{a} and P. T\"{o}rm\"{a} for helpful discussions and for sharing their data with us. IST,  BVS, and NVP acknowledge support from the Simons Foundation grant SFI-MPS-NFS-00006741-07 in the Simons Collaboration on New Frontiers in Superconductivity. BC, EK, BVS and NVP acknowledge support from EPSRC through Grant No. EP/X01245X/1. The DiagMC calculations were performed using King's Computational Research, Engineering and Technology Environment (CREATE). This work used the ARCHER2 UK National Supercomputing Service (https://www.archer2.ac.uk) \cite{Archer}.
\end{acknowledgments}

\clearpage
 \onecolumngrid
\section{End Matter}
\subsection{Bold4+ scheme} \label{Appendix:Bold4+}
\begin{figure*}[htbp]
\noindent
\includegraphics[scale=0.325]{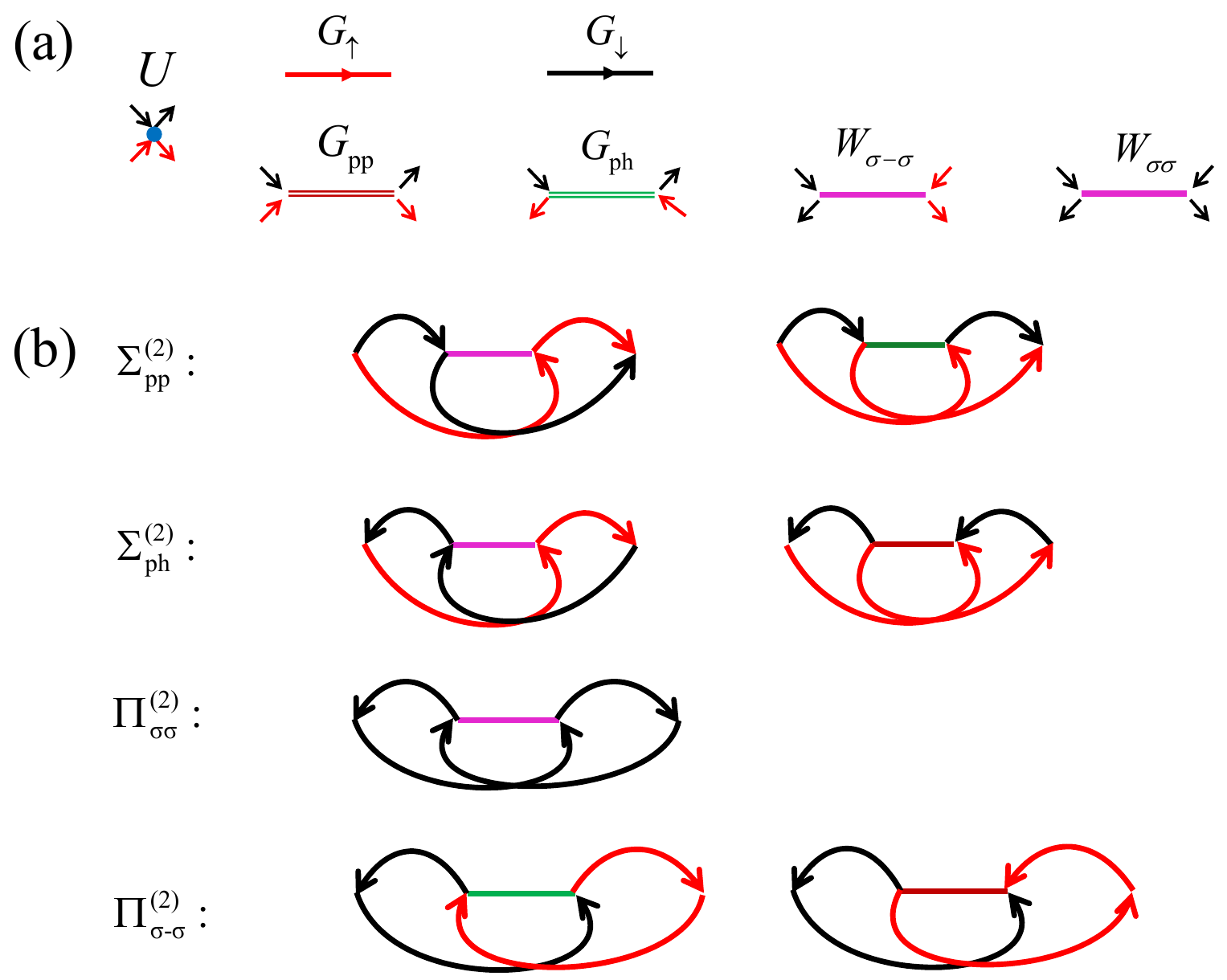}
\includegraphics[scale=0.325]{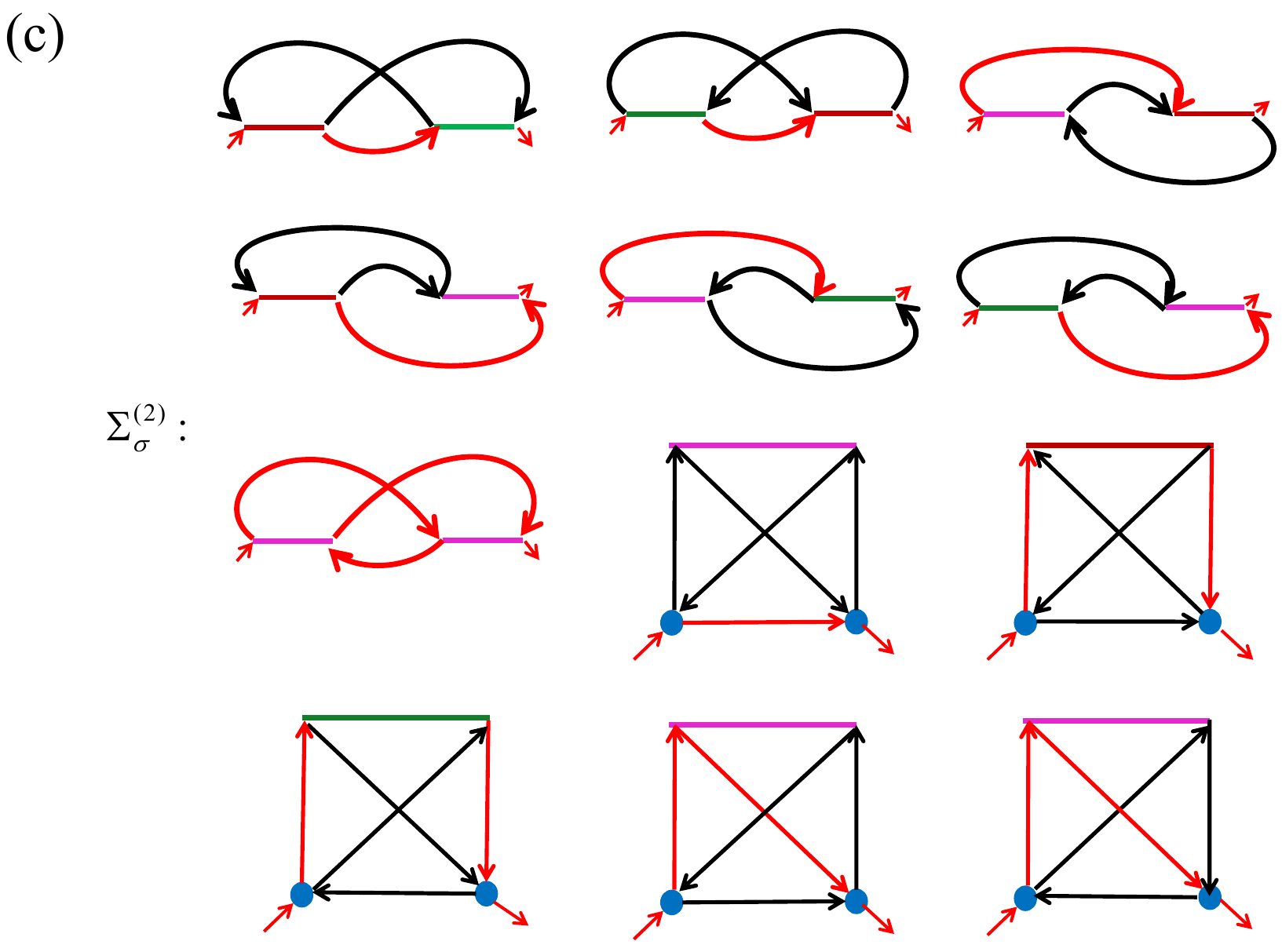}
\caption{
(a) Color scheme coding of Bold4+ diagrammatic elements.
(b) Leading-order vertex corrections to particle-particle, $\Sigma^{(2)}_{pp}$,
    particle-hole, $\Sigma^{(2)}_{ph}$,
    and screening, $\Pi^{(2)}_{\sigma \sigma'}$, proper self-energies.
(c) Leading-order vertex corrections to the single-particle proper self-energy
    $\Sigma^{(2)}_{\sigma}$.
}
\label{Fig5}
\end{figure*}

The Bold4+ scheme starts with solving a set of four self-consistent equations accounting for renormalization of single- and two-particle channels for contact interaction using the lowest-order diagrams \cite{MagCor2017}. More specifically, the four channels involved are the Green's function, $G_{\sigma}$, screened interaction, $W_{\sigma \sigma'}$, particle-hole propagator $G_{ph}$), and the particle-particle propagator $G_{pp}$. All propagators are functions of only one space-time variable (by definition, they start and end with the bare on-site interaction $U$, see panel (a) in Fig.~\ref{Fig5}) and are matrixes with respect to the atom labels. Self-energy diagrams for each propagator are defined in terms of the same propagators; they also start and end with $U$ leading to a set of self-consistent Dyson's equations; see Eqs.~(2)--(3) in Ref.~\cite{MagCor2017}.

The lowest-order self-consistent solution (abbreviated as Bold4 approximation) is subsequently used as the basis, similarly to the Hartree-Fock procedure, for setting up an expansion in the number of propagators and $U$ vertexes within the generic shifted action approach \cite{ShiftAct2016} when diagrams accounted for in the self-consistent solution serve as counter-terms.
By considering next-order diagrams on top of the Bold4 approximation (they are listed in panels (b) and (c) in Fig.~\ref{Fig5}) we obtain the Bold4+ scheme.
It accounts for all diagrammatic contributions to $\Sigma_{\sigma}$ up to $U^4$; higher-order diagrams are included only in the form of embedded geometric series within the Bold4 basis.

\subsection{Resummation of the series}
\label{Appendix:resum-S}

\begin{figure}[h]
\centering
\includegraphics[width=1.0\linewidth] {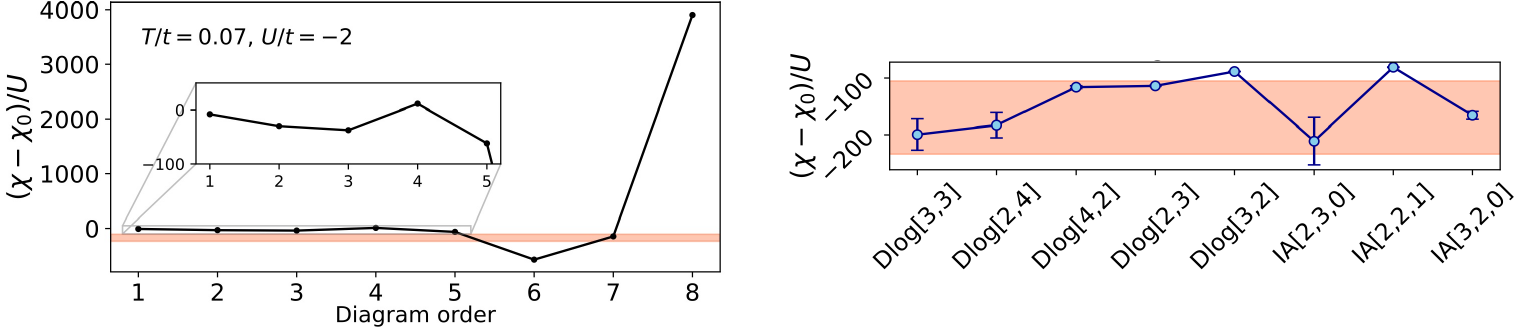}
\caption{Reconstruction of results from the diagrammatic series with controlled accuracy, demonstrated using data for the gapless case (i) close to $T_*=0.067(2)$ (see Fig.~\ref{fig:T_star}) at $T=0.07t$ and $U=-2t$. Left panel: the partial sums of the divergent series for $(\chi-\chi_0)/U$ versus the expansion order of $\chi$ up to $N=8$. Right panel: resummed values and statistical errors from individual approximants: Dlog-Pad\'e~\cite{Pade1961}, labelled ``Dlog$[M,K]$'', and Integral Approximants~\cite{Hunter1979IA}, labelled ``IA$[M,K,L]$'', with the free integer parameters $L,M,K$ constrained by $M+K+2=N$ and $L+M+K+3=N$, respectively (see, e.g., \cite{PadeSK2019} for details). Despite the divergence of the series, the different approximants yield consistent results, leading to a reliable estimate of the value and error represented by the shaded bands.
}
\label{fig:dlog}
\end{figure}

\begin{figure*}[htbp]
\centering
\includegraphics[width=0.245\linewidth]{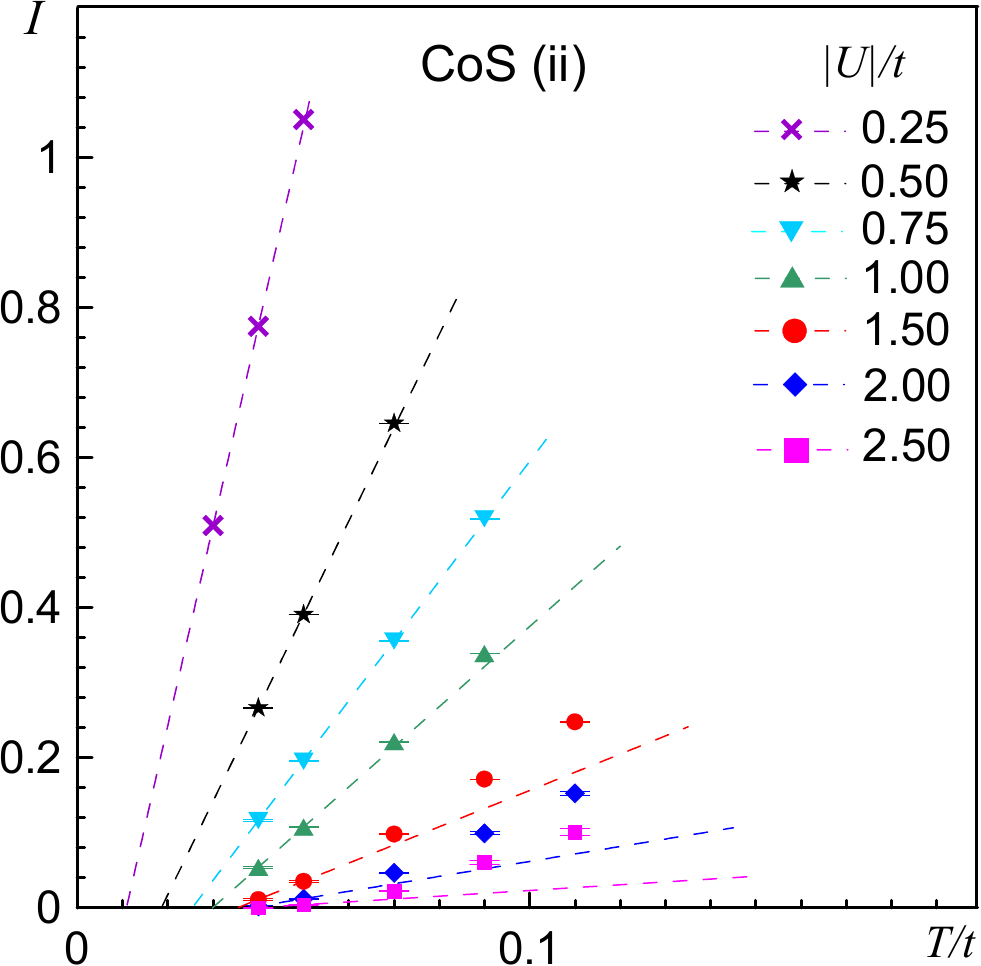}
\includegraphics[width=0.245\linewidth]{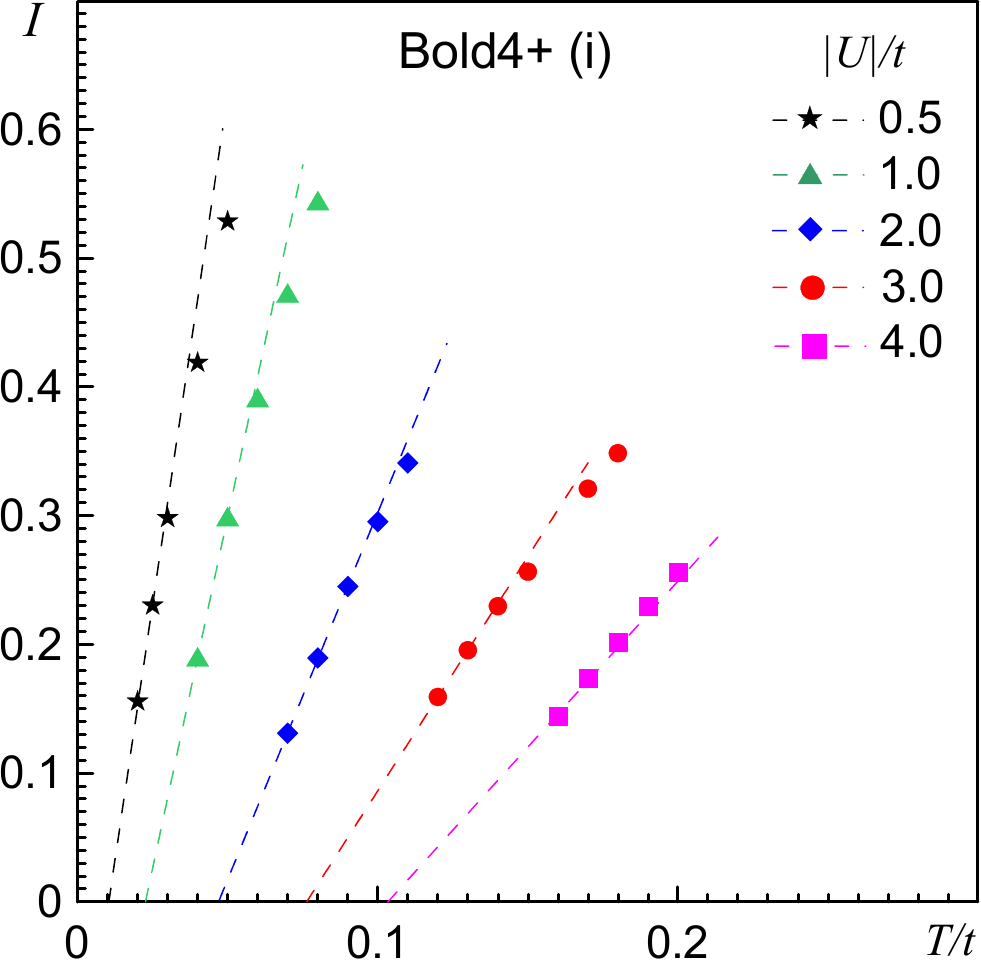}
\includegraphics[width=0.245\linewidth]{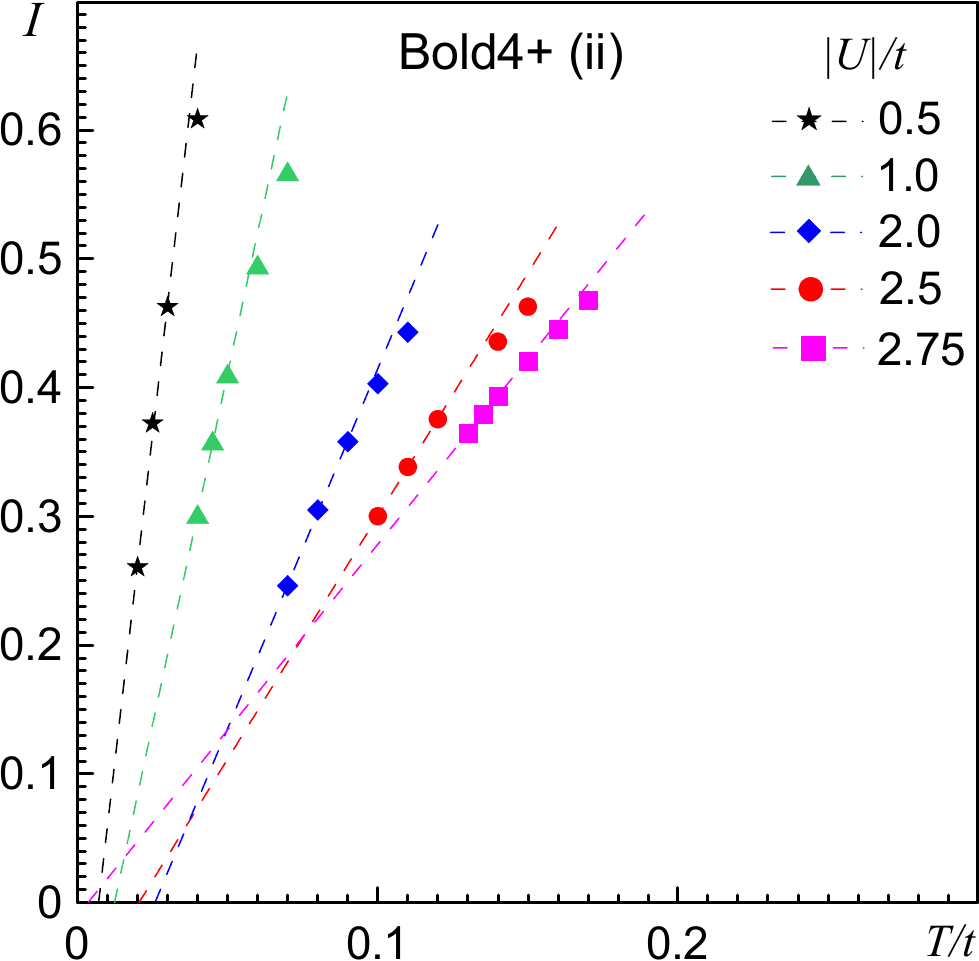}
\includegraphics[width=0.245\linewidth]{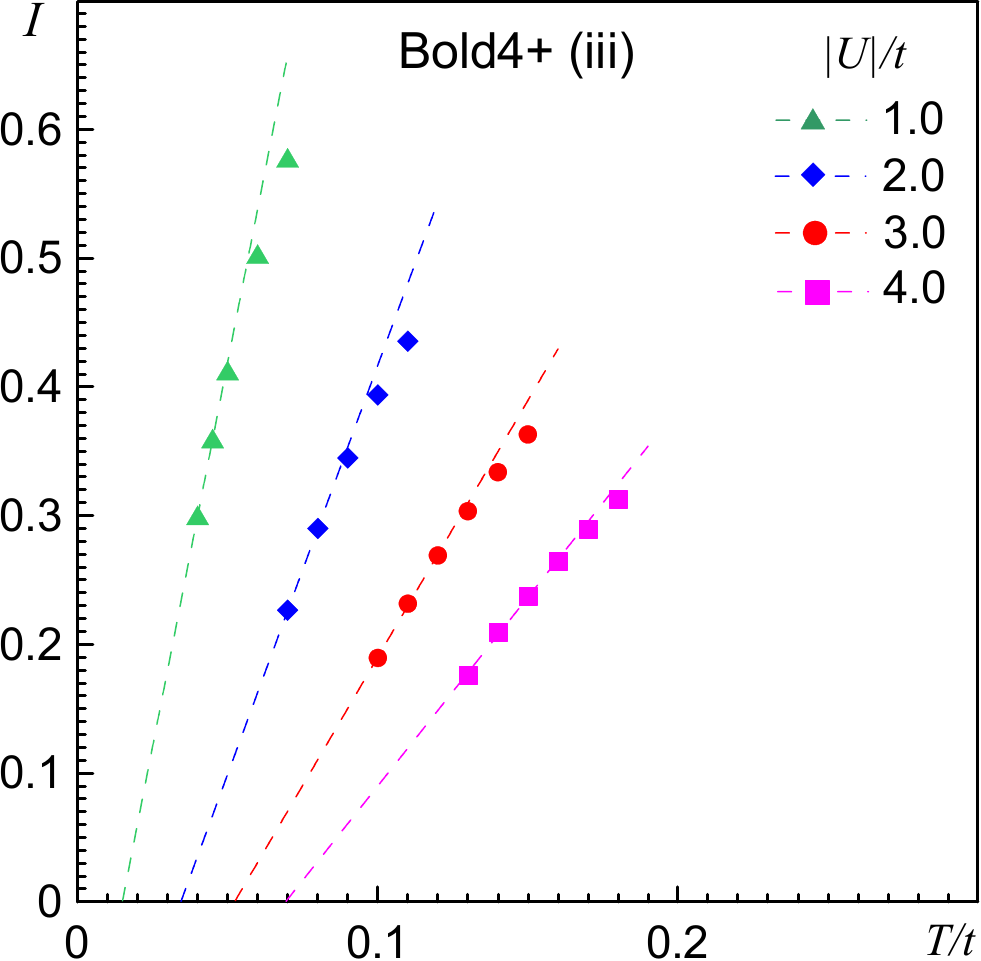}
\caption{
Extracting $T_c(U)$ from $I(T)$ from the CoS and Bold4+ data
(the same way as in Fig.~\ref{fig:straight}). 
Left to right: CoS for setup (ii); (2) Bold4+ for setup (i);
(3) Bold4+ for setup (ii); (4) Bold4+ for setup (iii). }
\label{I_T_data}
\end{figure*}

Starting from the diagrammatic expansion for the function $(\chi-\chi_0)$ in powers of $U$, the goal is to reconstruct the underlying function from the Taylor series coefficients alone. For this we employ the Dlog-Pad\'e and Integral approximant methods~\cite{PadeSK2019,Hunter1979IA}, which construct an approximating function whose Taylor series coefficients match exactly those of our diagrammatic expansion up to some maximal diagram order (8 in practice). The free parameters of the approximants are then varied to estimate the systematic error of the resummation. The approximants are parametrised as Dlog$[M,K]$ and IA$[M,K,L]$ with the free integer parameters $L,M,K$ constrained by $M+K+2=N$ and $M+K+L+3=N$, respectively, where $N$ is the number of coefficients used. In both cases, $K$ corresponds to the number of singularities possessed by the approximant, and $L$ corresponds to the order of the polynomial prefactor of the IA (see, e.g., \cite{PadeSK2019} for further details).
In practice, we restrict to ``balanced approximants'' where $M$ and $K$ are close to one another, which is a compromise between allowing a sufficient number of singularities while limiting the appearance of spurious singularities. All approximants that feature consistent singularity structures are used to determine the final answer, and the final combined error bar is obtained from resampling of the values from individual approximants within their errorbars. An example of such a resummation at $U/t = -2$ close to $T_*$ is shown in Fig.~\ref{fig:dlog}. The codes and data-analysis protocols used in this work were thoroughly cross-validated against existing results for the Hubbard model (as discussed in Ref.~\cite{Kozik2024}), the Bold4+ scheme at weak coupling, and
recent experimental data for the Hubbard model~\cite{Harvard_experiment2025}, as shown in Ref.~\cite{Currie2025}.

\vspace{-8mm}
\subsection{Inverse paring susceptibility $I(T)$}
\label{Appendix:I-T}

\vspace{-3mm}

\vspace{-8mm}
\subsection{Attractive Hubbard $U$ from electron-phonon interaction} \label{Appendix:electron-phonon}

Consider standard Holstein-type interaction Hamiltonian
with local (on-site) coupling
\be
H_{\rm e-ph} = \sum_{i} \,
\frac{g}{\sqrt{2\omega_0}} \,
\left[ n_i b_{i} + h.c. \right] .
\label{Holstein}
\ee
Here $n_{i}$ is the total electron density on site $i$
and $b_{i}$ is the annihilation operator of an optical mode with frequency $\omega_0$. According to this Hamiltonian,
energy gain for having two electrons on the same site,
$E_2 = - 2g^2/\omega_0^2 $, is larger than energy gain
for two electrons on separate sites, $2E_1 = - g^2/\omega_0^2$,
by an amount $U = -  \frac{g^2}{\omega^2_0}$.
This leads to an estimate $g^2= |U| \omega_0^2$ that is independent of the lattice type or band structure and
corresponds to moderate values of the dimensionless parameter $\lambda$ in the Migdal-Eliashberg theory for \textit{conventional superconductors with dispersive conduction bands}:
$\lambda = N_F \ (g^2/\omega^2_0) \to |U| N_F$
where $N_F$ is the density of states at the Fermi surface per spin component. For square lattice at small density
$N_F =1/4\pi t$, and this estimate corresponds to
$0.24 \lesssim \lambda \lesssim  0.32$  for
$3 \lesssim |U|/t \lesssim 4$.


\begin{thebibliography}{99}
%
\bibitem{Khodel1990} V.A. Khodel and V.R. Shaginyan,
Superfluidity in systems with condensed Fermi surface,
              JETP Lett., \textbf{51} (9), 553-555, (1990).
%
\bibitem{Kopnin2011} N.B. Kopnin, T.T. Heikkil\"{a}, and G.E. Volovik,
High-temperature surface superconductivity in topological flat-band systems,
              Physical Review B \textbf{83} (22), 220503 (2011).
%
\bibitem{Heikkila2011} T.T. Heikkil\"{a}, N.B. Kopnin, and G.E. Volovik,
Flat bands in topological media,
              JETP Lett. \textbf{94}, 233 (2011).
%
\bibitem{PT2015} S. Peotta and P. T\"{o}rm\"{a},
Superfluidity in topologically nontrivial flat bands,
              Nature Communications, \textbf{6} (1), 8944 (2015).
%
\bibitem{PT2016} A. Julku, S. Peotta, T.I. Vanhala, D.-H. Kim, and P. T\"{o}rm\"{a},
Geometric origin of superfluidity in the Lieb-lattice flat band,
              Phys. Rev. Lett. \textbf{117}, 045303 (2016).
%
\bibitem{PT2017} L. Liang, T.I. Vanhala, S. Peotta, T. Siro, A. Harju, and P. T\"{o}rm\"{a},
Band geometry, Berry curvature, and superfluid weight,
              Phys. Rev. B \textbf{95}, 024515 (2017).
%
\bibitem{Iskin2019} M. Iskin,
Origin of flat-band superfluidity on the Mielke checkerboard lattice,
              Phys. Rev. A \textbf{99}, 053608 (2019).
%
\bibitem{note8}
For flat bands with quadratic touching points (not considered here) one may also need to add a logarithmic factor $\ln t/|U|$ \cite{Iskin2019}.
%
\bibitem{QGT-Rev}  T. Liu, X.-B. Qiang, H.-Zh. Lu, and X.C. Xie,
Quantum geometry in condensed matter,
           National Science Review \textbf{12} (3), nwae334 (2025).
%
\bibitem{Yu2025}  J. Yu, B.A. Bernevig, R. Queiroz, E. Rossi, P. T\"{o}rm\"{a}, B.-J. Yang,
Quantum geometry in quantum materials,
            npj Quantum Materials \textbf{10}, 101 (2025).
%
\bibitem{PT2022-1} K.-E. Huhtinen, J. Herzog-Arbeitman, A. Chew, B.A. Bernevig, and P. T\"{o}rm\"{a},
Revisiting flat band superconductivity: dependence on minimal quantum metric and band touchings.
              Phys. Rev. B \textbf{106}, 014518 (2022).
%
\bibitem{PT2022-2} J. Herzog-Arbeitman, A. Chew, K.-E. Huhtinen, P. T\"{o}rm\"{a}, and B.A. Bernevig,
Many-body superconductivity in topological flat bands.
           Preprint at https://arxiv.org/abs/2209.00007 (2022).
%
\bibitem{multi2022}  P. T\"{o}rm\"{a}, S. Peotta, B.A.  Bernevig,
Superconductivity, superfluidity and quantum geometry in twisted multilayer systems,
           Nature Review Physics \textbf{4}, 528 (2022).
%
\bibitem{PT2025} R.P.S. Penttil\"{a} , K.-E. Huhtinen, and P. T\"{o}rm\"{a},
Flat-band ratio and quantum metric in the superconductivity of modified Lieb lattices,
           Commun. Phys. \textbf{8}, 50 (2025).
%
\bibitem{Berg2020}
J.S. Hofmann, E. Berg, and D. Chowdhury,
Superconductivity, pseudogap, and phase separation in topological flat bands: a quantum Monte Carlo study,
	Phys. Rev. B \textbf{102}, 201112 (2020).
%
\bibitem{Berg2023}
J.S. Hofmann, E. Berg, and D. Chowdhury,
Superconductivity, charge density wave, and supersolidity in flat bands with tunable quantum metric,
     Phys. Rev. Lett. \textbf{130}, 226001 (2023).
%
\bibitem{Lenihan2022} C. Lenihan, A.J. Kim, F. \v{S}imkovic IV, and E. Kozik,
Evaluating Second-Order Phase Transitions with Diagrammatic Monte Carlo: Néel Transition in the Doped Three-Dimensional Hubbard Model,
              Phys. Rev. Lett. \textbf{129}, 107202 (2022).
%
\bibitem{Currie2024} B. Currie and E. Kozik,
Fractional quantum Hall states by Feynman's diagrammatic expansion,
https://arxiv.org/abs/2412.21064 (2026).
%
\bibitem{Lieb1989} E.H. Lieb,
Two Theorems on the Hubbard Model,
              Phys. Rev. Lett. \textbf{62}, 1201 (1989).
%
\bibitem{KT}  J. M. Kosterlitz and D. J. Thouless,   Long-Range Order and Metastability in Two-Dimensional Solids and Superfluids, J. of Physics C: Solid State Physics {\bf 5}, L124 (1972); Ordering, Metastability and Phase Transitions in Two-Dimensional Systems, J. of Physics C: Solid State Physics
{\bf 6}, 1181 (1973).
%
\bibitem{Prokofev2007} N. Prokof'ev and B. Svistunov,
Bold Diagrammatic Monte Carlo Technique: When the Sign Problem Is Welcome,
              Phys. Rev. Lett. \textbf{99}, 250201 (2007).
%
\bibitem{VanHoucke2010} K. Van Houcke, E. Kozik, N. Prokof’ev, B. Svistunov,
Diagrammatic Monte Carlo,
              Physics Procedia \textbf{6}, 95 (2010).
%
\bibitem{Kozik2010} E. Kozik, K. Van Houcke, E. Gull, L. Pollet, N. Prokof'ev, B. Svistunov, and M. Troyer,
Diagrammatic Monte Carlo for Correlated Fermions,
              Euro Phys. Lett. \textbf{90}, 10004 (2010).
%
\bibitem{Kozik2024} E. Kozik,
Combinatorial summation of Feynman diagrams,
              Nature Communications, \textbf{15}, 7916 (2024).
%
\bibitem{Pade1961} G.A. Baker Jr,
Application of the Pad\'{e} approximant method to the investigation of some magnetic properties of the Ising model,
              Phys. Rev. \textbf{124}, 768 (1961).
%
\bibitem{Hunter1979IA} D. L. Hunter and G. A. Baker, Jr., Phys. Rev. \textbf{B} 19, 3808 (1979).
%
\bibitem{PadeSK2019} F. \v{S}imkovic IV and E. Kozik,
Determinant Monte Carlo for irreducible Feynman diagrams in the strongly correlated regime.
              Phys. Rev. B \textbf{100}, 121102(R) (2019).
%
\bibitem{MagCor2017} F. \v{S}imkovic IV, Y. Deng, N.V. Prokof'ev, B.V. Svistunov, I.S. Tupitsyn, and E. Kozik, Magnetic correlations in the two-dimensional repulsive Fermi-Hubbard model,
              Phys. Rev. B \textbf{96}, 081117(R) (2017).
%
\bibitem{NK} D.R. Nelson and J.M. Kosterlitz,
Universal Jump in the Superfluid Density of Two Dimensional Superfluids,
              Phys. Rev. Lett. \textbf{39}, 1201 (1977)
%
\bibitem{Electride2025} Ch. Ding, Y. Zhu, Q. Lu, Zh. Zhang, D. Shao, T. Huang, Y. Han, J. Wang, J. Sun,
Flat-Band Lieb Electride with Emergent quantum Phase Transitions and Superconductivity,
              Chinese Phys. Lett. \textbf{42}, 110709 (2025).
%
\bibitem{Archer} Beckett, G., Beech-Brandt, J., Leach, K., Payne, Z., Simpson, A., Smith, L., Turner, A., Whiting, A.
ARCHER2 Service Description. (2024)
%
\bibitem{ShiftAct2016} R. Rossi, F. Werner, N. Prokof'ev, and B. Svistunov,
Shifted-action expansion and applicability of dressed diagrammatic schemes,
              Phys. Rev. B \textbf{93}, 161102(R) (2016).
%
\bibitem{Harvard_experiment2025} M. Xu, L. H. Kendrick, A. Kale, Y. Gang, C. Feng, S. Zhang, A. W. Young, M. Lebrat, and M. Greiner, A neutral-atom Hubbard quantum simulator in the cryogenic regime, Nature \textbf{642}, 909 (2025).

\bibitem{Currie2025} B. Currie, J. Sturt, E. Kozik, Numerical validation of an ultracold Hubbard quantum simulator, https://arxiv.org/abs/2508.18041 (2025)

\end{thebibliography}
\end{document}